\documentclass[a4paper]{spie}  
\usepackage[]{graphicx}

\title{Dissecting Galaxies with Adaptive Optics}

\author{R.~Davies\supit{a}, 
H.~Engel\supit{a}, 
E.~Hicks\supit{b},
N.M.~F\"orster Schreiber\supit{a},
R.~Genzel\supit{a},
L.J.~Tacconi\supit{a},
F.~Eisenhauer\supit{a},
and
S.~Rabien\supit{a}
\skiplinehalf
\supit{a}
Max Planck Institut f\"ur extraterrestrische Physik, 
 Postfach 1312, Garching 85741, Germany\\
\supit{b}
Astronomy Department, University of Washington, 
Seattle, WA 98195-1580, USA
}

\authorinfo{Further author information (Send correspondence to
  R.D.)\\E-mail: davies@mpe.mpg.de}

\newcommand{\arcsec}{\hbox{$^{\prime\prime}$}}

\begin{document} 
  \maketitle 

\begin{abstract}
We describe several projects addressing the growth of galaxies and
massive black holes, for which adaptive optics is mandatory to reach
high spatial 
resolution but is also a challenge due to the lack of guide stars and
long integrations. 
In each case kinematics of the stars and gas, derived from integral
field spectroscopy, plays a key role. 
We explain why deconvolution is not an option, and that instead the
PSF is used to convolve a physical model to the required resolution. 
We discuss the level of detail with which the PSF needs to be known,
and the ways available to derive it. 
We explain how signal-to-noise can limit the resolution achievable and
show there are many science cases that require high, but not
necessarily diffraction limited, resolution.
Finally, we consider what requirements astrometry and photometry place on adaptive optics performance and design.
\end{abstract}



\section{The Growth of Galaxies and their Central Black Holes}
\label{sec:intro}

The existence of
tight relations between supermassive black holes (BH) and the stellar
velocity dispersion, luminosity, mass, and light concentration of
their host galaxies has been a major discovery of the last 
decade\cite{kor95,mag98,fer00,geb00,mar03,har04,gra07}.
These correlations, which imply that BHs and their host galaxies
co-evolve, have resulted in a widespread observational and theoretical
effort to understand their physical origin. 
In addition to the flow of gas inward from the host galaxy to the central
regions where it can fuel a BH, 
a plausible mechanism capable of linking the innermost
nuclear region to the whole galaxy is the feedback outward from this
accreting BH (an AGN) to the host galaxy.
Understanding the co-evolution of BHs and their host galaxies has become
one of the most important challenges of modern astrophysics.
In this section, we briefly look at 3 science topics that address this
issue from different perspectives.
For all of these cases, kinematics, derived from integral field
spectroscopy at high spatial resolution is a key element of the
analysis.

\subsection{Black Hole Masses in nearby Active Galaxies}
\label{sec:bhmass}

\begin{figure}[t]
\centering
\includegraphics[width=14cm]{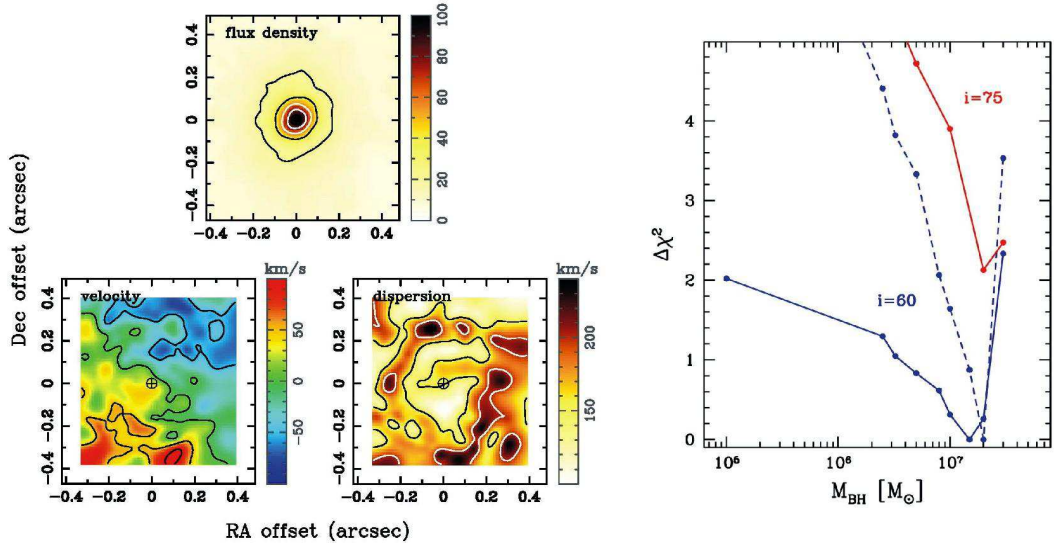}
\caption{Left: stellar continuum distribution and stellar kinematics
  in the central arcsec (80\,pc) of NGC\,3227. Right: increase
  in $\chi ^2$ from its minimum value for Schwarzschild models as a
  function of black hole mass. The results for 2 inclinations
  (60$^\circ$ and 75$^\circ$) are shown. 
The dashed line shows the impact of including rescaled Gauss
  Hermite terms $h_3$ and $h_4$. Adapted from Ref~\citenum{dav06}; see the electronic version for colour figures.
}
\label{fig:science1}
\end{figure}

Measuring BH masses is a fundamental step in deriving the relations above.
One of the best methods is to model the stellar kinematics, which can
only be done directly for nearby galaxies.
This is complementary to other methods such as reverberation mapping
which is based on temporal variability of the broad emission lines, and can not only verify such masses but in principle yields information
about the geometry of the broad line region.
NGC\,3227 was the first Seyfert~1 in which Schwarzschild orbit
superposition modelling was
applied to the stellar kinematics\cite{dav06}.
The result (see Fig.~\ref{fig:science1}) was a 1$\sigma$ range for the
BH mass of (0.7--2.0)$\times$10$^7$\,M$_\odot$.
This is a robust result, although the range is rather large because of
the limited signal-to-noise and also from degeneracies between the
mass of the various components in the model. 
It is a little lower than, although just consistent with, masses
derived from other methods based on reverberation mapping and X-ray
variability; and suggests that NGC\,3227 may lie a little below the 
$M_{\rm BH}-\sigma^*$ relation.
This is intriguing because NGC\,3227 is a barred galaxy, and there is
some evidence that barred galaxies as a class may preferentially lie
below the relation, perhaps hinting that the relations between black
holes and their host galaxies are more complex than first thought.

For this work, adaptive optics is essential to reach the highest
resolution possible in order to resolve the sphere of
influence of the black hole; integral field spectroscopy is
required to map the stellar kinematics in 2D, since the method can be
unreliable with the limited information available from longslit
spectroscopy; and a good estimate of the PSF is required for the modelling to ensure that beam smearing (see Sec.~\ref{sec:kin}) is treated correctly.

\subsection{Star Formation in Major Mergers}
\label{sec:merger}

\begin{figure}[t]
\centering
\includegraphics[width=12cm]{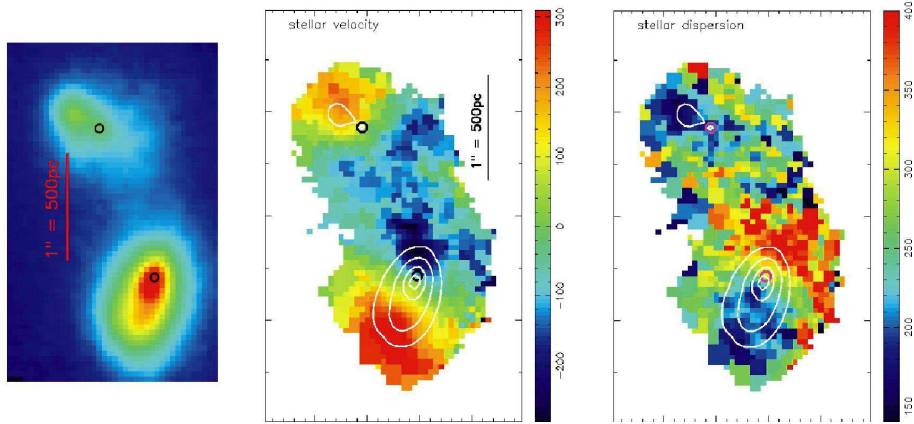}
\caption{Stellar continuum distribution and stellar kinematics for the
  prototypical merger NGC\,6240. The black hole
  locations\cite{max07} are indicated.
Adapted from Ref~\citenum{eng10}; see the electronic version for colour figures.
}
\label{fig:science2}
\end{figure}

Merging is one of the fundamental ways in which galaxies evolve and
grow.
Indeed, ULIRGs, the most luminous objects in the local universe,
appear to be predominantly mergers whose
prodigious infrared emission arises from dust heated by
obscured star formation activity.
At high redshift, mergers were common events and may have dominated
the cosmic star formation density at $z\sim2$, the epoch of
stellar mass assembly.
At the present day, however, mergers are rare.
There are therefore only a few nearby examples -- of which NGC\,6240
is one -- from which we can learn how the merging process proceeds in
detail.
Building on the existing qualitative picture of this
merger\cite{tec00}, 
we have
recently used integral field spectroscopy (see
Fig.~\ref{fig:science2}) and hydrodynamical
simulations to perform a quantitative analysis of the mass, luminosity,
age, and origin of its young and old stellar populations\cite{eng10}.
The mass-to-light ratios indicate the bulk of the stellar mass is
older than $\sim$1\,Gyr.
Coupled with other properties (size, dispersion, etc) of the nuclei, 
this strongly suggests they are the progenitor bulges.
Only 20--30\% of the bolometric luminosity, and
an even smaller fraction of the mass, originates in the recent
starburst.
The star formation history is an important issue, because it has a
significant impact on the starburst scale derived from the
observations.
We have found that the star formation rate associated with the recent
starburst has been
gradually increasing over the last $\sim$100\,Myr -- and will continue
to do so for another 100--300\,Myr, at which point, during the final
coalescence, the merger will become a ULIRG.

The distance of this prototypical merger is such that
$1\arcsec = 500$\,pc, and so adaptive optics is mandatory in order to
spatially resolve it.
Integral field spectroscopy is needed to provide a complete view of
the different geometries and kinematics for the molecular gas, ionised gas, and stars. 
In this case, to cover the whole system requires a field of 3\arcsec,
which SINFONI can only do with its intermediate 0.05\arcsec\ pixel
scale.
This means that our resolution is pixel limited -- and so the role of
adaptive optics was to provide enhanced rather than diffraction
limited resolution.
Independent of the resolution actually achieved, a good estimate of the PSF is required, as for the previous example, in order to treat beam smearing correctly when deriving the kinematic model.

\subsection{Galaxy Evolution at High Redshift}
\label{sec:highz}

\begin{figure}[t]
\centering
\includegraphics[width=14cm]{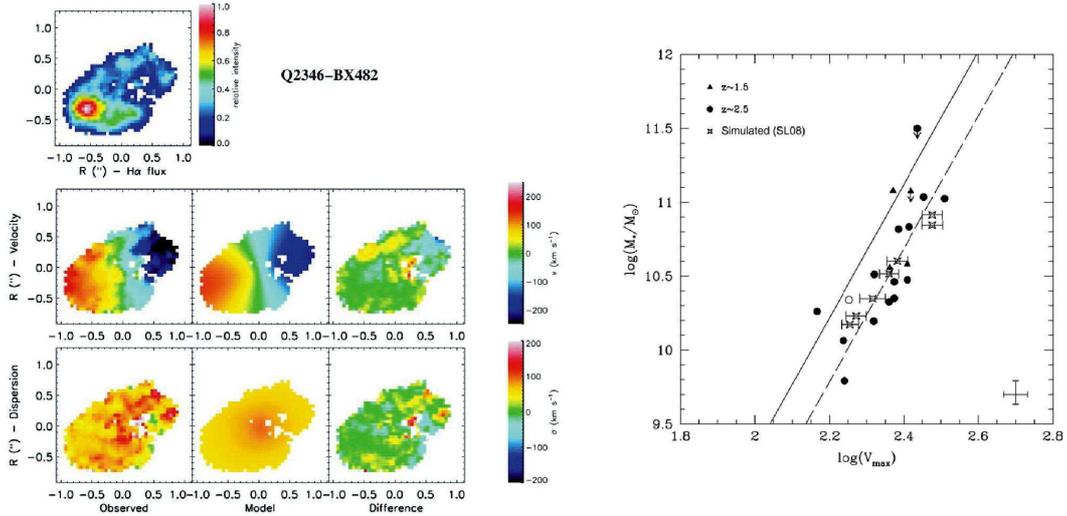}
\caption{Left: H$\alpha$ distribution of BX\,482 together with the
  observed and modelled kinematics. Right: Tully-Fisher Relation for
  disk galaxies at $z\sim2$, which are offset from the local $z=0$
  relation, indicating evolution in the scaling parameters.
Adapted from Ref.~\citenum{cre09}; see the electronic version for colour figures.
}
\label{fig:science3}
\end{figure}

The highest star formation rates and greatest incidence of QSO
activity occured at $z=2$--3, indicating that this was the time when
galaxies were evolving most rapidly.
Much effort has been put into studying galaxies at this epoch, and
most of what is known about them has come from work on the `deep
fields' which have been extensively surveyed at all accessible
wavelengths.
A recent addition to the toolbox are detailed integral field
observations, making use of adaptive optics where
possible\cite{for09}.
One of the key results from this work is that, although high-z
galaxies are clumpy, these clumps are often part of a massive gas-rich
star-forming disk -- i.e. they are not indicative of mergers.
This has led to a new perpsective in which, complementary to mergers,
smooth accretion of cold gas is now understood to play a major role in
galaxy evolution at early cosmic time.
The properties of the disks -- their sizes, rotation velocities, and
masses -- have enabled scaling relations, such as the Tully-Fisher
relation, to be traced out to $z\sim2$ and have shown that these
relations do evolve\cite{cre09}, as can be seen in Fig.~\ref{fig:science3}.
Measuring these properties requires an automated fitting procedure to
match a dynamical model to the 2D observations, taking into account
the PSF.

Adaptive optics is an important aspect of this work, since high-z
galaxies are at most 1--2\arcsec\ across.
However, without extremely long integrations, diffraction limited
observations are impractical because of the faint surface brightness
of the galaxies.
Thus, on 8-m class telescopes, a major role of AO is to provide seeing
enhancement, to achieve resolutions of $\sim$0.15\arcsec\
corresponding to about 1\,kpc.
A more crucial issue is the sky coverage: the fields from which the
galaxies are selected are designed to be free of bright stars.
There are very few targets with bright nearby stars, and 
even if a laser guide star (LGS) is used, tip-tilt correction is difficult.
Fortunately, the relaxed requirements on resolution lead directly to
relaxed requirements on tip-tilt;
and in some cases, it has been possible to observe without tip-tilt at
all.

Dispensing with tip-tilt completely\cite{dav08} yields 100\% sky
coverage, and is possible on the VLT because the fast (03\,Hz) guiding
provides a pseudo tip-tilt correction.
It opens the door to many important targets that would not otherwise
be observable with adaptive optics, with applications ranging from
young stellar objects, star clusters, nearby galaxies, and high
redshift galaxies.

\section{Recovering the PSF}
\label{sec:psf}

One crucial aspect of the science cases above is that all they
require knowledge of the PSF.
As we show below, although there are a number of ways for an observer
to estimate the PSF, they all have serious limitations.
Because of this, we would argue that PSF reconstruction based on
wavefront sensor (and perhaps ancilliary) data should, at
some level, be a requirement of all future AO systems.

In general, the methods available to derive it
depend strongly on the level of precision with which it must be known.
Searches for binaries or planets around bright stars represent the
extreme case where every detail is crucial in order to distinguish
between small faint objects and the PSF structure.
However, the requirements for many -- perhaps most -- science
applications are less demanding.
For extragalactic science it is usually sufficient to use a simple
combination of two analytical functions.
The PSF can generally be matched quite well by the sum of a narrow
Gaussian, to fit the core, and either a Moffat or another
Gaussian to trace the extended wings in the halo.
Such representations are ideal when one is fitting a simple model
(kinematic or morphological) to the science object in order to derive
its intrinsic properties.
Greater detail is unnecessary because the accuracy of the analysis is
limited by the model and/or signal-to-noise.
With a little invention, there are in principle a wide variety of methods
available to infer the shape of the PSF, as shown by the list
below.
However, we also note that all these methods have serious disadvantages
which limits their general applicability.
\vspace{-3mm}

\begin{description}

\item{\em Use an isolated reference star}\cite{now07,now08,now10}.
This is generally the method recommended to an observer but
there are a number of reasons why it is often unsuitable:
(i) it is inefficient due to the time needed to
observe a separate reference star sufficiently often;
(ii) because the PSF is observed at a different time, and for a much
shorter period, than the science target, there is no guarantee that
it is a reliable representation of the PSF in the science data;
(iii) a good PSF estimate requires that the intensity and distribution
of flux on the WFS is the same as the science target guide star.
If both are stars, this can work well;
but for AGN and other sources that are either extended or superimposed on a
bright or irregular background, it is simply unreliable;
(iv) if the wavefront reference is not the science target, one needs
to take account of anisoplanatism and find a pair of stars at the
appropriate separation,
one of which matches the guide star magnitude and the other
of which can be observed by the science camera.
With a laser guide star, the wavefront reference will be the same
regardless of the science target, and so point (iii) is resolved.
But the issues remain for (i), (ii) and, in the context of
tip-tilt stars, (iv).\vspace{-1mm}

\item{\em Extrapolate it from surrounding stars}.
Depending on the local environment close around the science target, it
may be possible to measure the PSF from nearby stars.
This in effect the same as the method used in stellar fields, but
there are also some extragalactic targets where it can be
applied\cite{pri04}.
More often, the stars will be distributed over a wide field and so it
will be necessary to account for anisoplanatism. 
Several methods have been developed to estimate an off-axis PSF; 
and in principle these could be inverted to derive an on-axis PSF
from off-axis stars.
Typically, they require knowledge of the $C_N^2$ distribution
through the atmosphere or observations of calibration frames
containing many stars.
But it is also possible to make a an
approximation to the way the PSF varies across a wider field using the
science data alone, as long as at least one or two stars or compact
objects are detected\cite{cre05,cre06}.
However, while this method may work for imaging, it cannot be used for
spectroscopy (either longslit or integral field) because the field of
view is simply too small.\vspace{-1mm}

\item{\em Extract it from the science data}\cite{dav04a,dav04b,dav06,dav07,neu07,fri10}. 
If there is any feature (either continuum or emission/absorption line)
in the science target that is spatially unresolved, it can be used as
an estimate of the PSF.
In nearby AGN, the broad line region is only a few lightdays across
and is always unresolved (in QSOs, its size can be measured in
light years but, because of their greater distance, is still unresolved).
Alternatively, the near-infrared non-stellar continuum associated with
AGN is only 1--2\,pc across and hence unresolved on 8-m class
telescopes at distances of $\sim20$\,Mpc or more.
The spatial distribution of both these quantities can be extracted
using the spectral information available in a datacube.
While this is extremely successful for spectrosopic studies of AGN, it
cannot be applied to most objects, nor to imaging.\vspace{-1mm}

\item{\em Derive it by comparison to other higher resolution
  data}\cite{mul06,eng10}.
This method works because convolution is associative.
The rationale is that convolution of the intrinsic source $S$ with a
PSF $P$ yields the observed source $O = S \otimes P$;
and that one can define a broadening function $F$ which, when convolved with
the higher resolution observation $O_h$, reproduces the lower
resolution observation $O_l = O_h \otimes F$.
Putting these together one can write 
$S \otimes P_l = S \otimes P_h \otimes F$
And so the lower resolution PSF is $P_l = P_h \otimes F$.
It is worth noting that if there is enough difference in
resolution, then $P_l \sim F$ and it is not necessary to know $P_h$
accurately.
While this can be highly successful, the
whole point of adaptive optics is to yield the highest resolution
data, making this method inapplicable.\vspace{-1mm}

\item{\em Reconstruct it from the wavefront sensor data}.
From the astronomer's perspective, this is an ideal option.
PSF reconstruction has been developed for both curvature
\cite{ver97} and Shack-Hartmann AO systems
\cite{cle08}.
It is undortunate that there are no general facilities for
reconstruction of the adaptive optics PSF at major observatories;
and that (to the best of our knowledge), the only current AO project that
specifically includes PSF reconstruction is ARGOS, the ground layer
LGS-AO system for the LBT\cite{rab10}.
Because of the importance of PSF knowledge to science analysis, we are
convinced that the effort spent in developing reconstruction algorithms would
yield significant benefits.

\end{description}

\section{Kinematics and (De-)Convolution}
\label{sec:kin}

Separating the PSF from the intrinsic structure in observed data is a
key part of most analyses.
This is the aim of deconvolution, and in some cases -- particularly
for high signal-to-noise data with a well-known PSF -- it can lead to
excellent results.
But it is an inverse problem and hence mathematically messy, 
tends to amplify noise, and can easily generate unreal artifacts such
as ringing.
And afterwards, there is still a PSF in the deconvolved data.
Although this new PSF is narrower, it is probably less well defined
and may vary with signal-to-noise across the field.
Whether one uses deconvolution depends on the aim of the analysis.
For an object with a lot of detail that cannot be
quantified in a convenient way -- for example structures on a
planetary surface -- then deconvolution may be the best (and only) way to
extract as much information from the data as possible.
Similarly, if all one needs to measure is an accurate position, then
deconvolution can help.

The alternative, the basis of galaxy fitting algorithms, is
to convolve an analytical model of the 
intrinsic structure with the PSF, compare the result to the
observations, and adjust the model iteratively.
The methods are complementary;
and the key difference, which enables them to be implemented in
mathematically contrasting ways, is
simply that this method is highly constrained.
For deconvolution, each pixel is an independent parameter, leading to
a huge number of degrees of freedom.
For convolution, only a few parameters are usually needed to
describe the analytical model.
This also means one can relatively easily estimate the
uncertainties in the model parameters.
And the method is highly suited to data with low signal-to-noise.
But perhaps the biggest advantage is that one ends up with an expression for
the intrinsic source properties.
However, this is also a limitation, because one only knows how well
that particular model matches the data.
The parameters derived may be biassed by more complex
structure that is not accounted for properly in a simple model, but
which the model still attempts to match.

\begin{figure}[t]
\centering
\hspace{6mm}
\includegraphics[width=3.7cm]{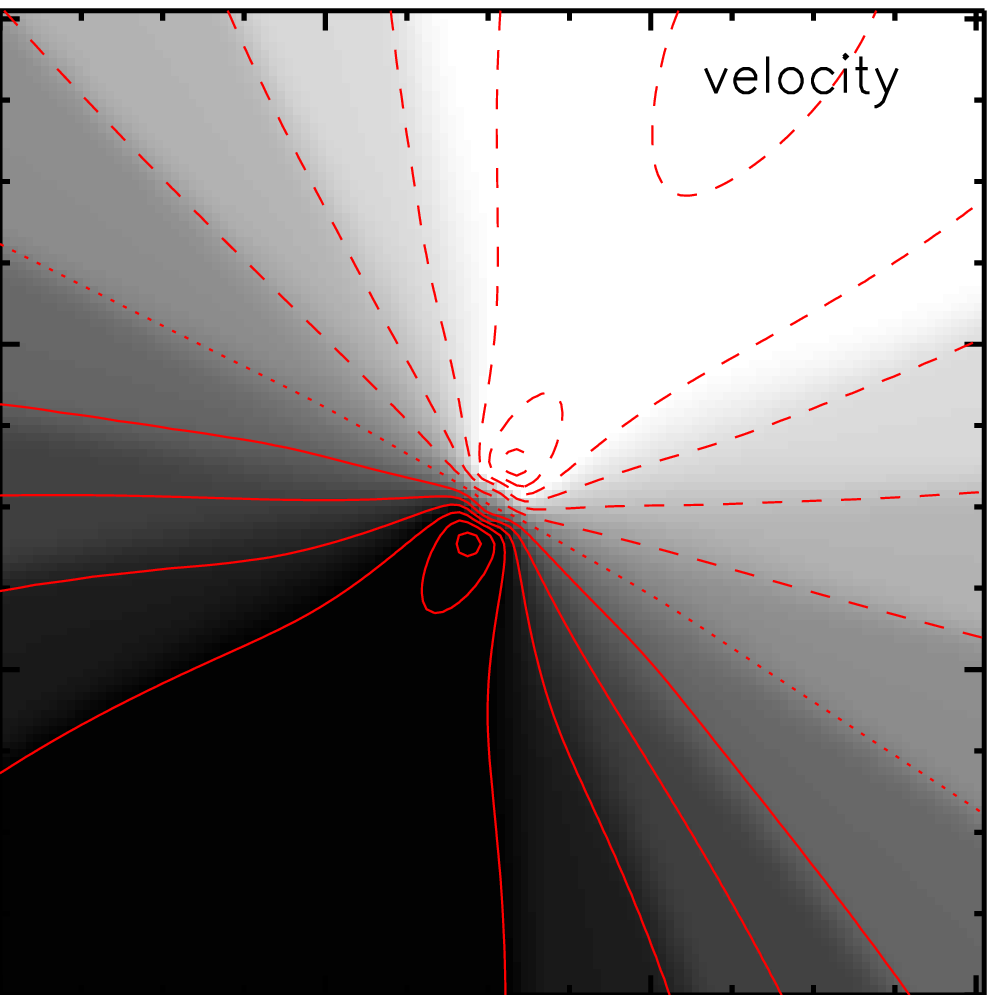}
\hspace{5mm}
\includegraphics[width=3.7cm]{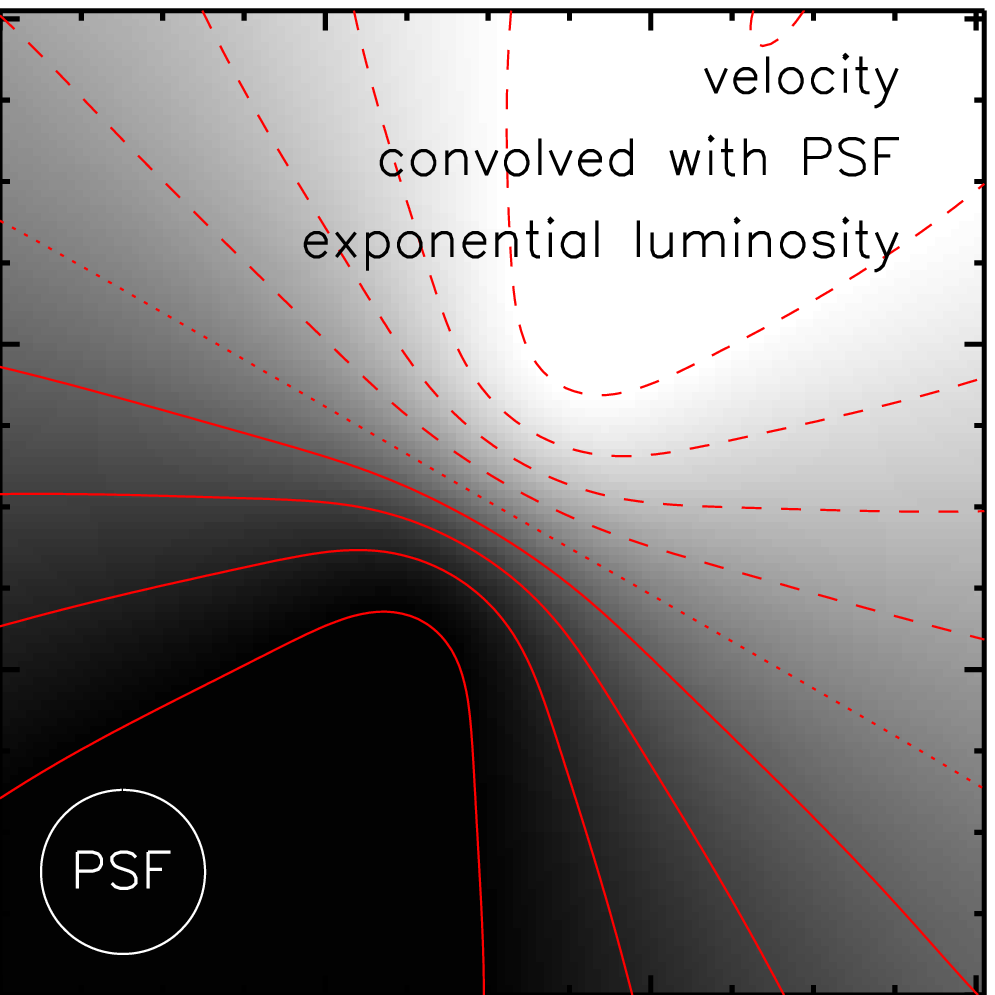}
\hspace{5mm}
\includegraphics[width=3.7cm]{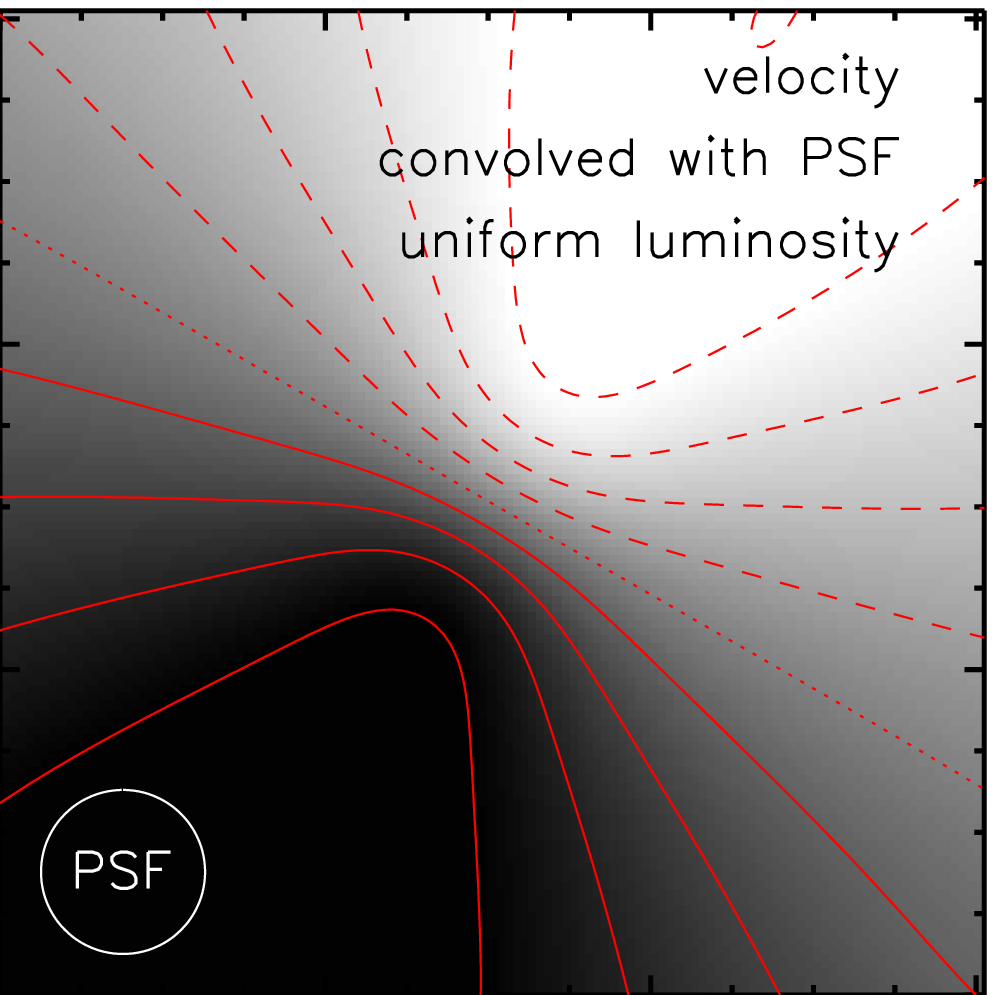}
\hspace{5mm}
\vspace{2mm}
\hspace{5mm}
\includegraphics[width=3.7cm]{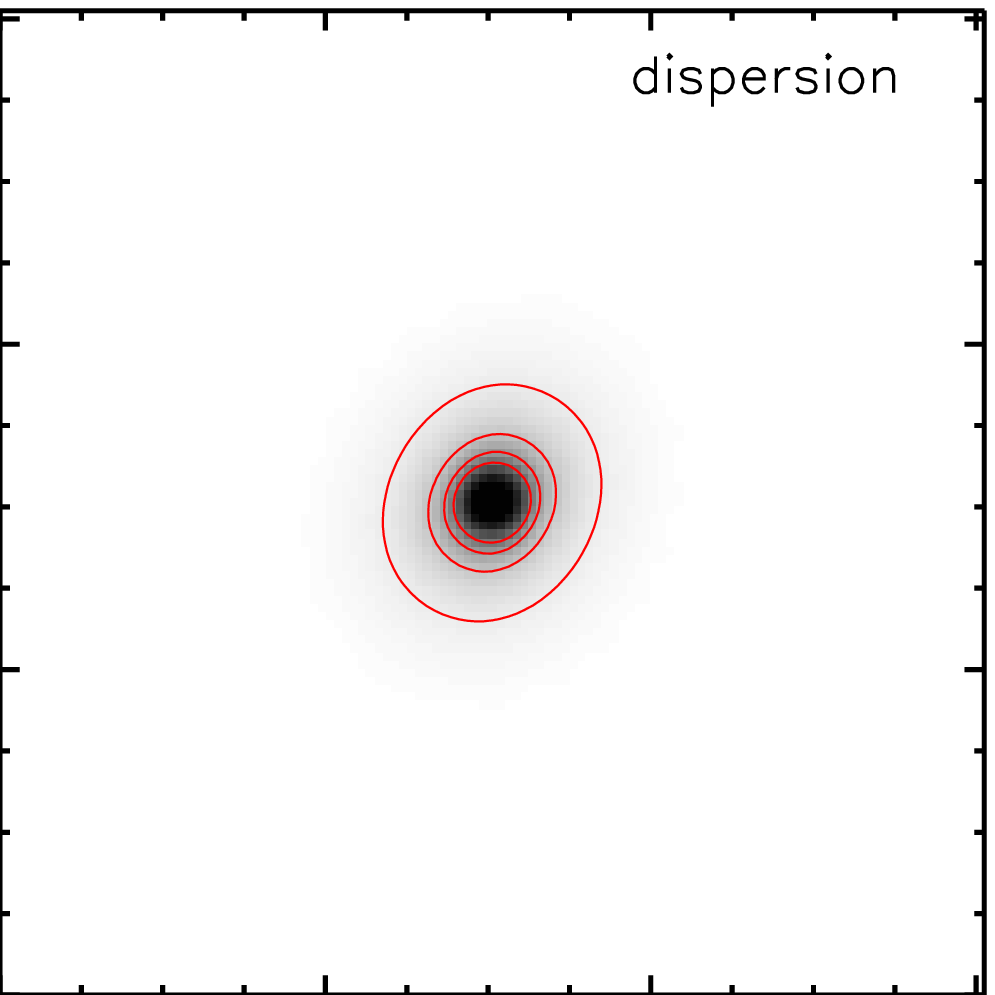}
\hspace{5mm}
\includegraphics[width=3.7cm]{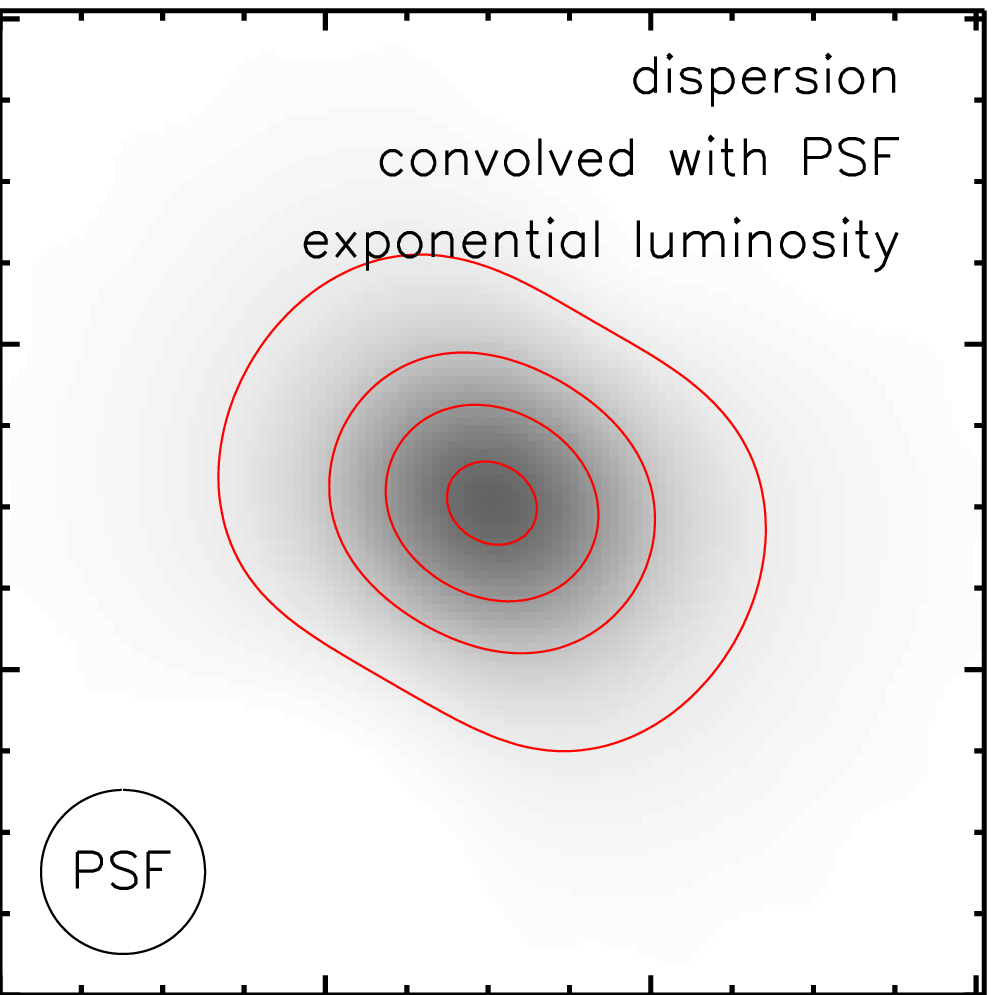}
\hspace{5mm}
\includegraphics[width=3.7cm]{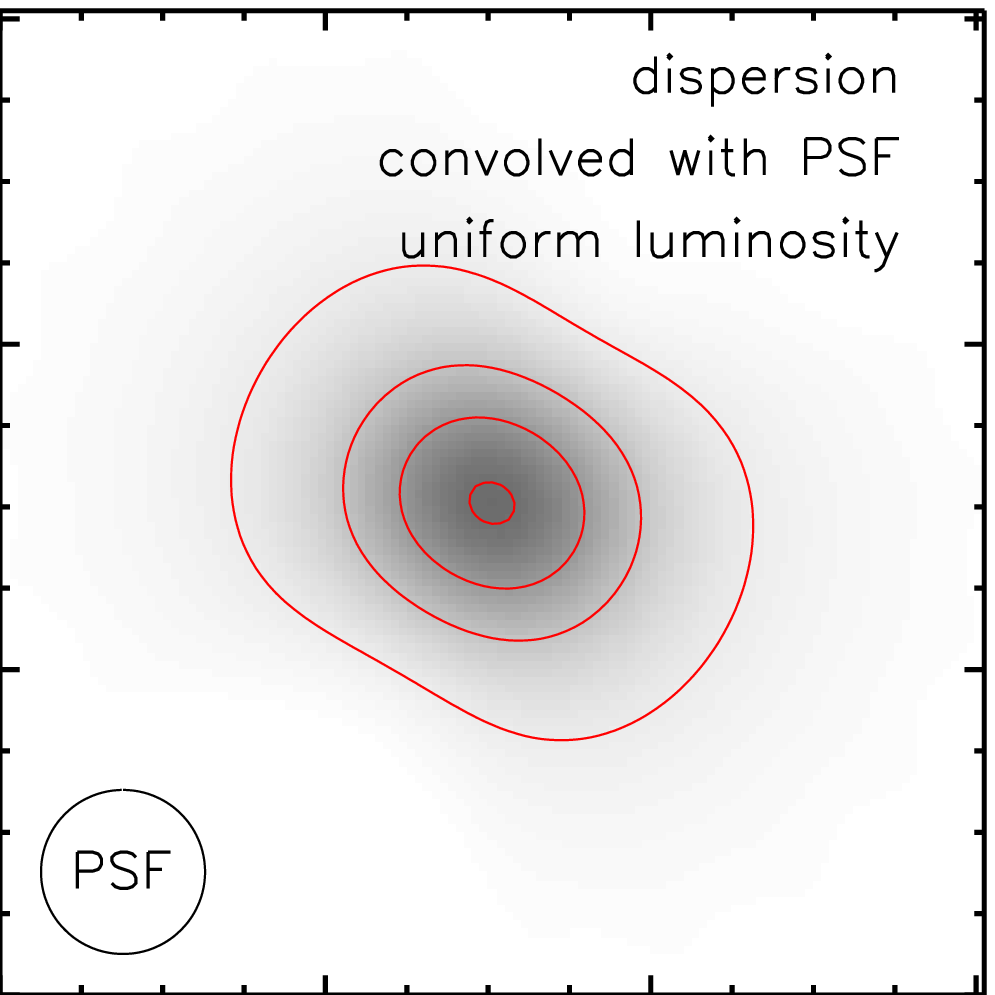}

\caption{Velocity and dispersion fields for matter in a
  self-gravitating disk around a supermassive black hole. In the field
  shown, the integrated mass of the disk is 5 times that of the black
  hole. The effect of PSF (beam smearing) on the kinematics is very
  dramatic. Due to cross-talk between these 2 quantities
  (i.e. velocity gradients on scales comparable to the PSF contribute
  to the dispersion), the 
  smeared kinematics cannot be deconvolved. The only option is to
  create a 3D kinematic model (2 spatial and 1 velocity dimension),
  convolve it with the PSF, and then extract the kinematics. By
  iterating one can constrain the model parameters.
The greyscales and contours are the same for all the velocity maps and
  all the disperison maps respectively.
}
\label{fig:kin}
\end{figure}

There are cases, such as kinematics, where deconvolution is
simply not an option due to the cross-talk between the velocity,
dispersion, and luminosity.
Figure~\ref{fig:kin} shows a model of an exponential disk around a
massive black hole, inclined at 35$^\circ$ from face-on.
The intrinsic kinematics (left panels) show a steep velocity gradient
and a high dispersion at the very centre.
The observed kinematics (centre panels) are dramatically different due
to the PSF.
The velocity gradient is shallower; and although the peak in
dispersion has gone, there is a larger area of high dispersion at an
orthogonal position angle.
The impact of the PSF depends also on the luminosity profile, since
beam smearing is always luminosity weighted.
The impact of having a uniform rather than exponential luminosity
profile (right panels) is more subtle. 
In this case it leads to a higher central dispersion by weighting
regions at the edge of the beam, that have
greater velocity differences, more highly.
All the science cases we have described in Section~\ref{sec:intro} --
and indeed one of the 
fundamental reasons to use integral field spectroscopy -- is to make
use of the 2D kinematics from emission or absorption features.
And it is for this reason that convolution with a model is the
only practical way to extract the required information from the data.

\section{Astrometry}
\label{sec:astrom}

\subsection{Black Hole Masses and Galaxy Evolution from Proper
  Motions}

Proper motions are a different way of measuring kinematics, and the
Galactic Centre is an example where this method has been applied very
successfully.
The full 3D orbits of several tens of stars are now known, enabling
precise measurements to be made of the mass and distance of the
central massive black hole. 
Astrometry will be an important capability on future ELTs and open
the way to measurements of the masses of putative intermediate mass
(i.e. $10^3$--$10^5$\,M$_\odot$) black holes which may exist at the
centres of globular clusters.
Such measurements can be made by tracking the relative proper motions
of thousands of stars within each cluster, and 
they will populate the very low mass end of the 
$M_{\rm BH}-\sigma^*$ relation.
Whether or not they lie on the relation, there is no doubt that they
will have a significant impact on our understanding of it.
The proper motions of the globular clusters themselves are also
important, since they enable the cluster orbits around the Galaxy
to be calculated.
The number of times a cluster passes close to, or through, the Galactic
plane will have a big impact on its evolution; and the existence of
kinematic families of clusters would imply distinct episodes of
cluster formation, leading to a more complete picture of the formation
and evolution of the Galaxy.
Astrometry can also play an important role in understanding the role
of dark matter and merging in galaxy evolution.
Dwarf Spheroidals in the Galactic Halo are dominated by dark matter,
and measuring their mass profiles (which requires astrometry in order
to break the degeneracy between this and anisotropy) will enable us to
test models of structure formation in the context of galaxy haloes.

\subsection{Astrometry Requirements on Adaptive Optics}

The main requirement is for the astrometric precision to be high and
uniform across a large field.
A 30--60\arcsec\ field that can provide positions of stars
across the bulk of a globular cluster, implies that multi-conjugate
adaptive optics is the most efficient method.
Experience from the Galactic Centre shows that with an 8-m class
telescope it is possible to achieve a relative astrometric precision
of $\sim$200\,$\mu$as in the H-band\cite{fritz10}, 
corresponding to about 0.5\% of the FWHM of the PSF.
If this performance is projected onto the E-ELT, one can hope to reach
a precision of about 40\,$\mu$as.
This would enable proper motions as small as 10\,$\mu$as\,yr$^{-1}$ --
equivalent to 5\,km\,s$^{-1}$ at 100\,kpc -- to be measurable within a
few years.
A study\cite{tri10}, related to the specific case of
MICADO\cite{dav10a,dav10b}, has shown that the various sources of
error can be controlled if appropriate measures are taken in the
instrument design and calibration.
However, it is worth noting that several of the sources of error are
directly related to adaptive optics. 
These are outlined in the list below, together with possible ways in
which their impact might be mitigated.\vspace{-3mm}

\begin{description}

\item{\em Instrumental and atmospheric effects on multiple NGS}.
MCAO makes use of several NGS.
Their relative locations on the focal plane will be affected by
a number of systematic effects:
variations in the plate scale from the telescope and the accuracy of
the derotation, as well as achromatic (i.e. dependent on airmass) and
chromatic (i.e. dependent on the star colour) differential refraction.
If the AO system tries to account for these relative motions, it will
warp the scientific field.
Since there are only a few stars, this may be only a low order
distortion, in which it can be corrected with post-processing using a
few reference objects in the field.
Alternatively, it should be possible to track the barycentre of each
NGS on a relatively slow timescale and hence compensate for these
effects directly within the AO system.
Malfunctioning actuators on the DM may also distort the science field;
however the impact of this has not yet been studied.
\vspace{-1mm}

\item{\em Differential tilt jitter}.
This stochastic effect is a serious issue for single-conjugate
adaptive optics. 
It dominates the anisoplanatism and, in long exposures, leads to the
well known PSF elongation as one moves further from the guide star.
It integrates down as $t^{-1/2}$, but on a 42-m ELT would require of
order 4\,hrs to reach an acceptable level for astrometry.
With MCAO, the impact is vastly reduced, but still requires
integrations of order 30\,mins to achieve 10\,$\mu$as errors.
To avoid saturation, individual exposures are likely to be no more
than a few seconds. Methods have been proposed to combine these
in such a way as to make use of the spatial correlations, and reduce
differential jitter to acceptable levels on significantly faster
timescales\cite{cam09}.\vspace{-1mm}

\item{\em PSF variations}.
In deep exposures, temporal variations in the PSF will average out.
On the other hand, spatial variations will not.
Anisoplanatism is dominated by residual jitter, which has already been
discussed. 
But high-order residuals will also subtly change the PSF shape as a
function of position and may lead to astrometric errors.
Using simulations from MAORY\cite{dio10} for LGS-MCAO, we have estimated the
impact, and concluded that -- based on these simulations -- it is
below $\sim$10\,$\mu$as and hence negligible.

\end{description}

\section{Photometry}
\label{sec:phot}

\subsection{Black Hole and Galaxy Evolution from Photometry}

QSOs represent a crucial piece of the puzzle about the co-evolution of
black holes and their host galaxies.
Studying them at high redshift will provide insights about the 
$M_{\rm  BH}-\sigma^*$ relation at early cosmic times.
However, measuring the properties of the host galaxy requires one to
subtract the central QSO, which is often several to tens of times
brighter -- a task that is increasingly difficult at higher redshifts,
and requires a good estimate of the PSF\cite{jah04}. 

An alternative way to study galaxy evolution is through relic
populations in nearby galaxies\cite{tol09}.
Accurate photometry of spatially resolved stellar populations enables
one to create colour-magnitude diagrams in order to derive a galaxy's
star formation history. 
The ultimate goal of such work is to apply these methods to galaxies
in the Virgo 
cluster which, at a distance of $\sim$17\,Mpc, is the nearest large
cluster, with over 2000 member galaxies of all morphological types.
Directly counting the number of stars on different branches enables
one to derive a coarse but robust star formation history for a galaxy
that, by probing the oldest stars with ages $>10^{10}$\,Gyr, can be
traced back to the early universe\cite{gre02}.
In the Leo Group (at a distance of $\sim$11\,Mpc), an imaging camera
on an ELT should comfortably reach the Horizontal Branch.
In closer galaxies, such as Cen\,A at $\sim$3\,Mpc, it will be possible
to measure the entire Red Giant Branch (RGB) down to the level of the
main sequence turnoff, and hence to address the problem of the
age-metallicity degeneracy.
Although sensitivity is important in such work, the crucial aspect is
resolution since this allows one to probe ever more crowded fields.
Working at the diffraction limit of an ELT means that in the Virgo
cluster, rather than probing the fringes
of galaxies at 3--4\,R$_{\rm eff}$, one can study the centres of
galaxies at $\sim$1\,R$_{\rm eff}$ where the bulk of the stellar mass
is located.

\subsection{Photometry Requirements on Adaptive Optics}

For both of the cases above, the uncertainties on the scientific result
depend directly on the accuracy with which the PSF is known.
For a QSO that is 5 times brighter than the host
galaxy, an error of only 2\% in the estimation of the Strehl ratio
leads to an uncertainty of about 10\% in the host galaxy magnitude.
And if the error on the Strehl is as much as 20\%, then one might not
even detect the host galaxy at all.
For stellar photometry, achieving a precision of 0.03\,mag means
measuring the flux in a PSF to within 3\%, which requires knowing the
profile out to a relatively large radius.

From the astronomer's perspective, reconstruction of the on-axis PSF
from WFS data would be the ideal solution.
Yet good PSF reconstruction, specifically when one or more LGS are
used in the wavefront sensing, remains a challenge.
In this respect a large field of view is an asset, regardless of
whether the science target fills it, because
it becomes highly likely that suitable empirical PSFs can be
found within the field even at high galactic latitudes.
If the PSFs do not vary much across the field, they can be combined
and used directly.
This is the essence of photometry packages, which successfully derive
the PSF from the science data itself even in crowded fields.
One aspect that needs to be assessed further is the impact on
photometric accuracy of spatial variations across the large field.
Simulations show that, even with MCAO, if these are not taken into
account, the photometric scatter increases markedly.
A possible interim solution, that has been used with SCAO data, is
to split the field into smaller sections across which the PSF is
approximately uniform.

Alternatively if one can parametrize several PSFs in different
locations it may be possible to determine how each parameter varies
across the field and hence interpolate the PSF at any position.
A positive step in addressing this by developing a simple model
for the PSF has been taken by the MAORY consortium \cite{dio10}.

\section{Conclusions}

We have looked at a number of science cases that
have a common goal of understanding the evolution of galaxies and
black holes through cosmic time.
They employ various techniques of integral field spectroscopy,
astrometry, and photometry.
Each science case benefits tremendously from -- and is often not
possible without -- adaptive optics.
Historically, adaptive optics systems have been designed to provide
the highest strehl ratios. However, many other issues play a role.
The question is then: in order to have the greatest scientific yield,
what is the best performance metric?\vspace{-3mm}

\begin{description}

\item{\em Strehl Ratio}.
While a high strehl ratio is always preferable, setting it as a performance
metric directly conflicts with extragalactic science, where guide
stars or tip-tilt stars are typically faint or sparse.\vspace{-1mm}

\item{\em Spatial Resolution}.
Good spatial resolution is mandatory for the science, but reaching the
diffraction limit is sometimes not possible due to limitations of the
instrumentation (e.g. pixel scale vs field of view) or
signal-to-noise.
Therefore, although resolution is important in some cases (e.g. photometry of
resolved stellar populations), it should not be a primary performance
criterion.\vspace{-1mm}

\item{\em Encircled Energy}.
This metric embodies the essence of both strehl ratio and resolution,
and can be an excellent indicator of performance in terms of the
contrast of the high resolution part of the PSF.
However, it is often specified within a relatively large diameter.
We propose it should be used as a criterion in terms of the fractional
energy within the core of the PSF without specifying what the core
width should be. This gives some freedom to optimise also other
critera, and in some cases even to dispense with a tip-tilt
star\cite{dav08}.\vspace{-1mm}

\item{\em PSF uniformity}.
All the science cases require knowledge of the PSF. While the PSF
can sometimes be estimated, it remains a major problem when analysing
data from current AO systems.
An important criterion is therefore that an AO system should enable
the PSF to be derived.
How this is done is less critical: a uniform PSF across a large field
(i.e. MCAO) means it can be measured directly; a parametrization would
allow it to be interpolated even if it varies within the field; or
ancilliary data might mean there is a way to reconstruct it.\vspace{-1mm}

\item{\em Sky Coverage}.
This is perhaps the most important criterion for extragalactic, and in
particular for high-z, work.
Statistical estimates of the fraction of sources that can be observed
at a given galactic latitude do not reflect how widely an AO system can
be used.
Instead, progress in astrophysics relies heavily on specific targets
and specific fields that are intensively studied in many different and
complementary ways.
An AO system has to be able to provide a good correction on these.
The scientific penality paid for insufficient sky coverage is
enormous.\vspace{-1mm}

\end{description}

\noindent
Our conclusion is that for future AO systems to have a major impact
on extragalactic science, they should:\newline
\indent(i) yield a high fraction of the flux in the PSF core (without specifically optimising strehl or resolution);\newline
\indent(ii) do so with maximum sky coverage (i.e. be optimised to
work at the faint limit); and\newline
\indent(iii) enable the PSF to be derived (either measured,
interpolated, or reconstructed).



\end{document}